\documentclass{aa}

\input{psfig}

\def\specchar#1{{\sc #1}}       
\def\CI{\mbox{C\,\specchar{i}}}

\def\OV{\mbox{O\,\specchar{v}}}

\def\OVIII{\mbox{O\,\specchar{viii}}}
\def\OVIII{\mbox{O\,\specchar{viii}}}
\def\FeXXI{\mbox{Fe\,\specchar{xxi}}}
\def\FeXXV{\mbox{Fe\,\specchar{xxv}}}
\def\NeIX{\mbox{Ne\,\specchar{ix}}}
\def\MgXI{\mbox{Mg\,\specchar{xi}}}
\def\SiXIII{\mbox{Si\,\specchar{xiii}}}
\def\SXV{\mbox{S\,\specchar{xv}}}
\def \SAIT #1 #2 {{\em Mem.\ Soc.\ Astron.\ It.\/} {\bf #1}, #2}
\def \MESS #1 #2 {{\em The Messenger\/} {\bf #1}, #2}
\def \ASTRNACH #1 #2 {{\em Astron. Nach.\/} {\bf #1}, #2}
\def \AAP #1 #2 {{\em Astron. Astrophys.\/} {\bf #1}, #2}
\def \AAL #1 #2 {{\em Astron. Astrophys. Lett.\/} {\bf #1}, L#2}
\def \AAR #1 #2 {{\em Astron. Astrophys. Rev.\/} {\bf #1}, #2}
\def \AAS #1 #2 {{\em Astron. Astrophys. Suppl. Ser.\/} {\bf #1}, #2}
\def \AJ #1 #2 {{\em Astron. J.\/} {\bf #1}, #2}
\def \ANNREV #1 #2 {{\em Ann. Rev. Astron. Astrophys.\/} {\bf #1}, #2}
\def \APJ #1 #2 {{\em APJ \/} {\bf #1}, #2}
\def \APJL #1 #2 {{\em Astrophys. J. Lett.\/} {\bf #1}, L#2}
\def \APJS #1 #2 {{\em Astrophys. J. Suppl.\/} {\bf #1}, #2}
\def \APSS #1 #2 {{\em Astrophys. Space Sci.\/} {\bf #1}, #2}
\def \ASR #1 #2 {{\em Adv. Space Res.\/} {\bf #1}, #2}
\def \BAIC #1 #2 {{\em Bull. Astron. Inst. Czechosl.\/} {\bf #1}, #2}
\def \JSQRT #1 #2 {{\em J. Quant. Spectrosc. Radiat. Transfer\/} {\bf #1}, #2}
\def \MN #1 #2 {{\em Mon. Not. R. Astr. Soc.\/} {\bf #1}, #2}
\def \MEM #1 #2 {{\em Mem. R. Astr. Soc.\/} {\bf #1}, #2}
\def \PLR #1 #2 {{\em Phys. Lett. Rev.\/} {\bf #1}, #2}
\def \PASJ #1 #2 {{\em Publ. Astron. Soc. Japan\/} {\bf #1}, #2}
\def \PASP #1 #2 {{\em Publ. Astr. Soc. Pacific\/} {\bf #1}, #2}
\def \NAT #1 #2 {{\em Nature\/} {\bf #1}, #2}
\def \SP #1 #2 {{\em Solar Phys.\/} {\bf #1}, #2}

\begin{document}

   \thesaurus{02.08.1; 
              06.03.2; 
              06.06.3; 
              06.20.1; 
              06.21.1; 
              06.24.1; 
             } 

   \title{Coronal loop hydrodynamics}
     \subtitle{The solar flare observed  
           on November 12 1980 revisited: the UV line emission.}
   \author{R.M. Betta \inst{1} 
           G. Peres \inst{2} 
           F. Reale \inst{2} 
           \and 
           S. Serio \inst{1,2}
          }

\institute{Osservatorio Astronomico di Palermo,
           Palazzo dei Normanni, 
           P.zza del Parlamento, 1,
           I-90134 Palermo, Italy 
           \and
           Dipartimento di Scienze Fisiche e Astronomiche,
           Sezione di Astronomia,
           University of Palermo,
           Palazzo dei Normanni, P.zza del Parlamento, 1
           I-90134 Palermo, Italy;  
           e-mail:peres@oapa.astropa.unipa.it
          }

\date{Received 2001 / Accepted }

\offprints{G. Peres \\e-mail:peres@oapa.astropa.unipa.it}

\maketitle

\begin{abstract}
We revisit a well-studied solar flare whose X-ray emission originating
from a simple loop structure was observed by most of the instruments on
board SMM on November 12 1980.  The X-ray emission of this flare, as
observed with the XRP, was successfully modeled previously.  Here we
include a detailed modeling of the transition region and
we compare the hydrodynamic results with the UVSP observations in two EUV
lines, measured in areas smaller than the XRP rasters, covering only
some portions of the flaring loop (the top and the foot-points).
The single loop hydrodynamic model, which fits well the 
evolution of coronal lines (those observed with the XRP and the 
\FeXXI\ 1354.1 \AA\ line observed with the UVSP)
fails to model the flux level and evolution of the 
\OV\ 1371.3 \AA\ line.


\end{abstract}

\section{Introduction}

Solar flares are very complex phenomena. They emit in a wide range
of wavelengths, including radio, optical, UV and X-rays.  
They involve many physical effects such as, for instance, 
chromospheric evaporation,
magnetic field reconnection and material ejection. 

There have been many examples of hydrodynamic models of flaring plasma
confined in solar coronal loops: for a list of such models and some
recent developments see \cite{Peres2000}.  In this context \cite{P87}
modeled the X-ray emission of the well-studied solar flare, which
occurred on November 12 1980 at 17:00UT.  In particular, they used the
Palermo-Harvard hydrodynamic loop model (\cite{P82}, thereafter
PH), to give an interpretation of the phenomena involved in this
event.  The modeled light curves were successfully compared with those
observed in some X-ray lines with the XRP on board SMM in a raster
region covering the flaring coronal loop.

The numerical resolution in the transition region (TR), however, was
not sufficient for a proper comparison with the EUV observations.  Most
EUV lines, in fact, are formed at temperatures below $10^6
\mathrm{K}$.  Using the new version of the PH code (\cite{betta97}),
having appropriate spatial resolution, here we revisit the modeling of
this well-studied flare.  In section 2 we describe the hydrodynamic
loop model; in section 3 we introduce our new simulations; in section 4
we compare the PH code numerical calculations with the observed line
light curves; in section 5 we summarize our conclusions.

\section{The flaring loop model}

The PH code (\cite{P82}) solves the one-fluid time-dependent
differential equations of mass, momentum and energy conservation for
the plasma confined in a solar magnetic loop, using a finite-difference
numerical scheme.  The model is one-dimensional, since the plasma is
confined inside a semicircular flux tube where bulk motion and heat
flow occur only along the magnetic field lines.  The present version of
the PH code (\cite{betta97}) can solve an asymmetric loop (e.g.
\cite{R2000}), however \cite{MacNeiceetal85} inferred that the flare
X-ray emission of the Nov 12 1980 flare was symmetric respect to the
apex, and therefore we limit to model half a loop.  We also assume
ionization and thermal equilibrium between ions and electrons and 
that the flux tube keeps its geometric shape.  The equations are
solved with proper accuracy along the loop and during the entire flare
evolution, since a regridding algorithm adapts the spatial grid
whenever and wherever needed.  The adaptive grid allows to reach very
high spatial resolutions (as small as $1 \mathrm{m}$) using a variable
number of grid points ($\sim 300-600$).

We consider the same loop parameters used in \cite{P87}, which were
inferred from an accurate analysis of the H$\alpha$ and X-ray images. 

The flare event is simulated by switching on a very strong impulsive
heating ${\cal Q}$.
We use the formulation 
$$  {\cal Q}(s,t)=H_{s}+H_0~f(t)~h(s) $$
where $s$ is the coordinate
parallel to the magnetic field lines, $t$ is time,
$H_s$, uniform and constant, balances ordinary coronal losses, 
while the second term in the right handside
represents the impulsive heating. This term may be due
to current dissipation.
Its spatial and temporal dependencies are separated into two factors. In
general we allow for various functional forms for $f(t)$ and $h(s)$.
Here we assume that the spatial term $h(s)$ has a Gaussian form of half-width
$\sigma$ and center $s_0$; 
$f(t)$ is a constant function for the first 180s and then decays exponentially
with an e-folding time of 60s. 

We repeat the following simulations: 
\begin{itemize}
  \item[A)]  heating applied at the top of the loop;
  \item[B)]  heating applied in a small region near the foot-points.
\end{itemize}
We compute the line emission from the temperature and density of the
plasma
along the loop obtained with the PH code, using the software ASAP written
by \cite{ASAP} in IDL.

\section{Numerical results}
The hydrodynamics of the coronal plasma subject to a strong flaring
heating has already been described extensively in previous papers 
in the literature.
Here we dwell on the most important details. 

\begin{figure}[htb]
\vspace{0.5cm}
\hspace{0cm}{\psfig{figure=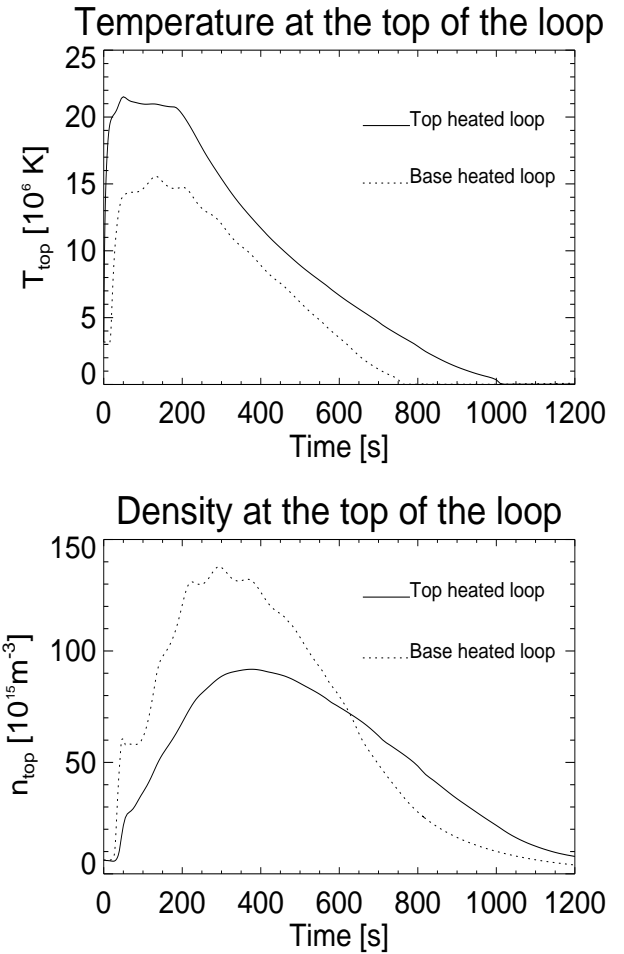,width=8.8cm,height=13cm}}
\caption{Temporal evolution of the temperature and hydrogen 
density at the top of the loop for the two models.}
\label{fig:td_evol} 
\end{figure}

The evolution of the temperature and the density at the top of
the loop are shown in Fig. \ref{fig:td_evol}.
In the first minute of evolution,
the coronal temperature increases from roughly 3 to 21 $~\mathrm{MK}$, 
for the model in which the energy is applied at the loop top
(simulation A), and to around $15~\mathrm{MK}$ for the model in which
the energy is applied at the base of the TR (simulation B).
During the early flare we find the largest differences
between the two models.

In both simulations A and B, the coronal density increases while
the impulsive heating is on.
After 200s of the simulations, the hydrogen density at the top of the
loop is of the order of $10^{17} \mathrm{m^{-3}}$.
In the pre-flare conditions the plasma hydrogen density at the top of
the loop was instead $\sim 7~10^{15} \mathrm{m^{-3}}$.

As the coronal temperature increases, the conductive front propagates
through the TR (the latter at the same time becomes steeper and
steeper), and the chromospheric plasma warms up and, after $\sim 15$s,
expands rapidly toward the corona ("chromospheric evaporation").

In simulation A, velocities at temperatures above $10^6 \mathrm{K}$ are
always toward the loop top after the first 24s and until the impulsive
heating is active, i.e until 360s.  In $\sim 30$s since the beginning
of this simulation, the plasma velocity is $\sim 300
\mathrm{km~s^{-1}}$ in the corona.  On the other hand, when the heating
is applied at the base of the loop (simulation B), the plasma is pushed
up since the very beginning and reaches velocities $\sim 400
\mathrm{km~s^{-1}}$ after 20s.

In both simulations a descendant flow of material occurs in the
chromosphere, below the region being ablated and evaporated; there the
plasma pressure is much lower than the pressure of the plasma evaporated
in the corona and the material is shocked and compressed toward the
bottom; velocities have much lower values than coronal ones. This
process is generally called "chromospheric condensation" (\cite{Fisher87}).

As the plasma starts to evaporate, in both simulations the plasma
temperature progressively increases along the whole loop; the emission
measure at any temperature increases (as well as the EUV flux).  The TR
moves along the loop toward the foot-points during the whole heating
phase.  After 240s, when the impulsive heating has decreased to around
1/3 of its maximum value, the corona starts to cool and the TR moves
gradually back to its initial position.

The simulations have been run to describe the plasma evolution for many
minutes after the end of the heating.  The loop cools and the coronal
density decreases until a thermal instability (\cite{Field65}) occurs:
then the whole loop cools very rapidly and reaches typical
chromospheric values ($\sim 10^4 \mathrm{K}$).  As a uniform and
constant heating term ($H_{s}$) is present in the equations, the whole
loop returns approximately to its stationary pre-flare equilibrium
after $\sim 3000$s.  The entire cycle - until the whole coronal
atmosphere returns exactly to the hydrostatic pre-flare conditions and
the velocities decrease to negligible values - takes a few hours.

\section{Light curves}

This flare has been described by
\cite{MacNeiceetal85} and \cite{CP88} (CP88, thereafter).
We now summarize the most important features for a better comprehension
of what is discussed in the following.

It started at 17:00 UT and lasted less than 20 minutes. 
Four of the five instruments on board SMM registered the event: 
the Hard X-ray Burst Spectrometer (HXRBS), the Hard X-ray Imaging Spectrometer
(HXIS), the X-ray Polychromator (XRP) and the Ultra-Violet Spectrometer
and Polarimeter (UVSP).
We concentrate on the observations made with the XRP (\cite{xrp}) 
and with the UVSP (\cite{UVSP}).
In \cite{P87} the light curves from the whole raster in the X-ray lines
observed with the XRP were compared with the results of 
the previous version of the PH code. An extensive analysis
and interpretation of the hot component of this flare as well as of the spatial
heating distribution and evolution was also included in that paper.   
During this flare the FCS instrument of the XRP 
registered spectroheliograms in six different resonance
lines of some H-like ions or He-like: \OVIII\ 18.97 \AA\ ($T \sim 3~
10^{6} \mathrm{K}$); \NeIX\  13.45 \AA\ ($T \sim 4~ 10^{6} \mathrm{K}$);
\MgXI\ 9.17 \AA\ ($T \sim 6.5~10^6 \mathrm{K}$); \SiXIII\  6.65 \AA\ ($T
\sim 10^7 \mathrm{K}$); \SXV\  5.04 \AA\ ($T \sim 1.55~10^7
\mathrm{K}$); \FeXXV\  1.85 \AA\ ($T \sim 7~10^7 \mathrm{K}$).
These lines covered most of the coronal temperature range.

There are no qualitative differences between the light curves synthesized
with the new and the previous version of the PH code, except the
disappearance of the numerical noise which affected the results at the
lowest coronal temperatures.
Therefore we confirm that the previous simulations
carried out by \cite{P87} are adequate 
to model the X-ray line light curves.

Here we focus most of our attention on the UVSP data. 
This spectrometer had spectral resolution
1150-3600 \AA\ and spatial resolution better than $2 \mathrm{arcsec}$.
The raster range was $256 \times 256 \mathrm{arcsec^2}$.
The UVSP collected images of this event with a field of view of
$120 \times 120 \mathrm{arcsec^2}$ and pixels of $10 \times 10
\mathrm{arcsec^2}$ (i.e. $\sim 7250 \times 7250 \mathrm{km^2}$ on
the Sun).
Observations continued until the emitted flux increased considerably;
since that moment ($\sim$17:00:06 UT), rasters (a complete raster lasted
$\sim 15$s) were alternated to spectral scans in two different lines,
which lasted 30s, with spectral resolution 0.3{\AA}.
These two lines, simultaneously recorded in two separate
detectors, are
\begin{itemize}
 \item \FeXXI\ 1354.1 \AA\ line, formed at typical flaring plasma
       temperatures ($T\sim 10^{7} \mathrm{K}$);
 \item \OV\ 1371.3 \AA\ line, formed at TR temperatures 
       ($T\sim 2.2~10^{5} \mathrm{K}$).
\end{itemize}
Both lines are partially blended with nearby lines (\cite{Feldmanetal77}).
The \FeXXI\ 1354.1 \AA\ line partially blends with the
chromospheric \CI\ 1355.844 \AA\ line; anyway this \CI\ line is
very narrow ($\sim 0.12$ \AA), and, moreover, \cite{Chengetal79}
showed that, during the flare maximum and decay phase, the intensity
of the \CI\ 1355.844 \AA\ line was $\sim 20$\% of the \FeXXI\ 1354.1
\AA\ line.
The \OV\ 1371.3 \AA\ line instead appeared blended with a very narrow
line at 1371.37 \AA\ (not yet identified) during many flares observed
by Skylab, which was
not present in the quiescent phase (\cite{Feldmanetal77}).

The UVSP observed the rising phase in both lines.
The flux in the \OV\ 1371.3 \AA\ line, already high in the pre-flare
phase, peaked at 17:02, i.e. at the same time as the peak observed in
the hard X-rays. 
Then the emission in this line gradually decreased while the emission in
the \FeXXI\ 1354.1 \AA\ line was still increasing (CP88).
The light curve in the \OV\ 1371.3 \AA\ line was different from the other
ones observed simultaneously, because it peaked twice in the pre-flare
phase (at 16:58UT and at 16:59UT) and, moreover, the flux was higher where
a filament was observed in the H$\alpha$ images. \cite{MacNeiceetal85}
wrote that material expulsions appeared in this filament, at the velocity
of $60 \mathrm{km~s^{-1}}$ before the rising phase of the flare.
This filament disappeared after a short time and its disruption might
be correlated to the increased emission in the \OV\ 1371.3 \AA\ line.

\begin{figure}[htb]
\vspace{0.5cm}
\centerline{\psfig{figure=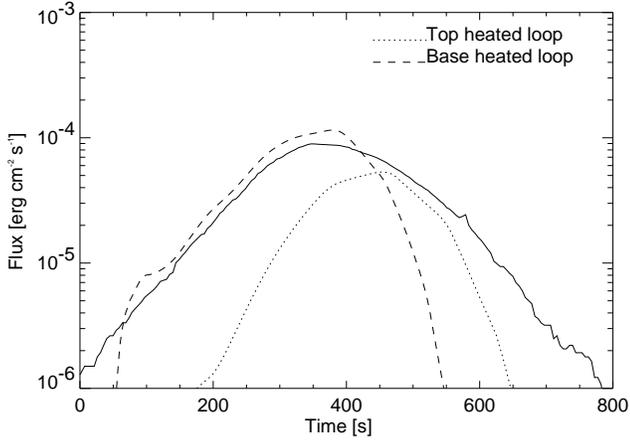,width=8.8cm}}
\caption{Light curve observed in the \FeXXI\ 1354.1 \AA\ line 
by the UVSP/SMM in kernel L1 (solid line) from CP88. Also plotted
are the light curves synthesized from hydrodynamic models with the 
emissivity tables computed by Monsignori Fossi 
(1994, private communication). 
In order to allow a proper comparison,
the emission is integrated around the top of the loop.}
\label{fig:l1fe} 
\end{figure}

\begin{figure}[htb]
\vspace{0.5cm}
\centerline{\psfig{figure=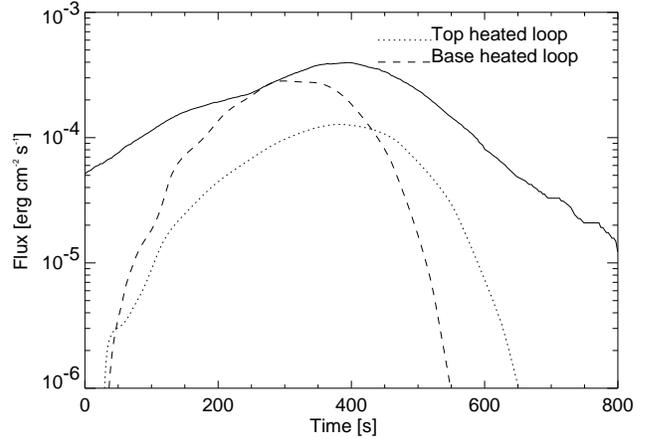,width=8.8cm}}
\caption{Light curve in the \FeXXI\ 1354.1 \AA\ line observed
by the UVSP/SMM in the whole raster (solid line) from CP88. 
Also plotted are the light curves synthesized with our code and with 
the 
emissivity tables computed by Monsignori Fossi 
(1994, private communication) for both models.
The line emission is integrated along the whole loop. 
}
\label{fig:rastfe} 
\end{figure}

\begin{figure}[tbp]
\vspace{0.5cm}
\centerline{\psfig{figure=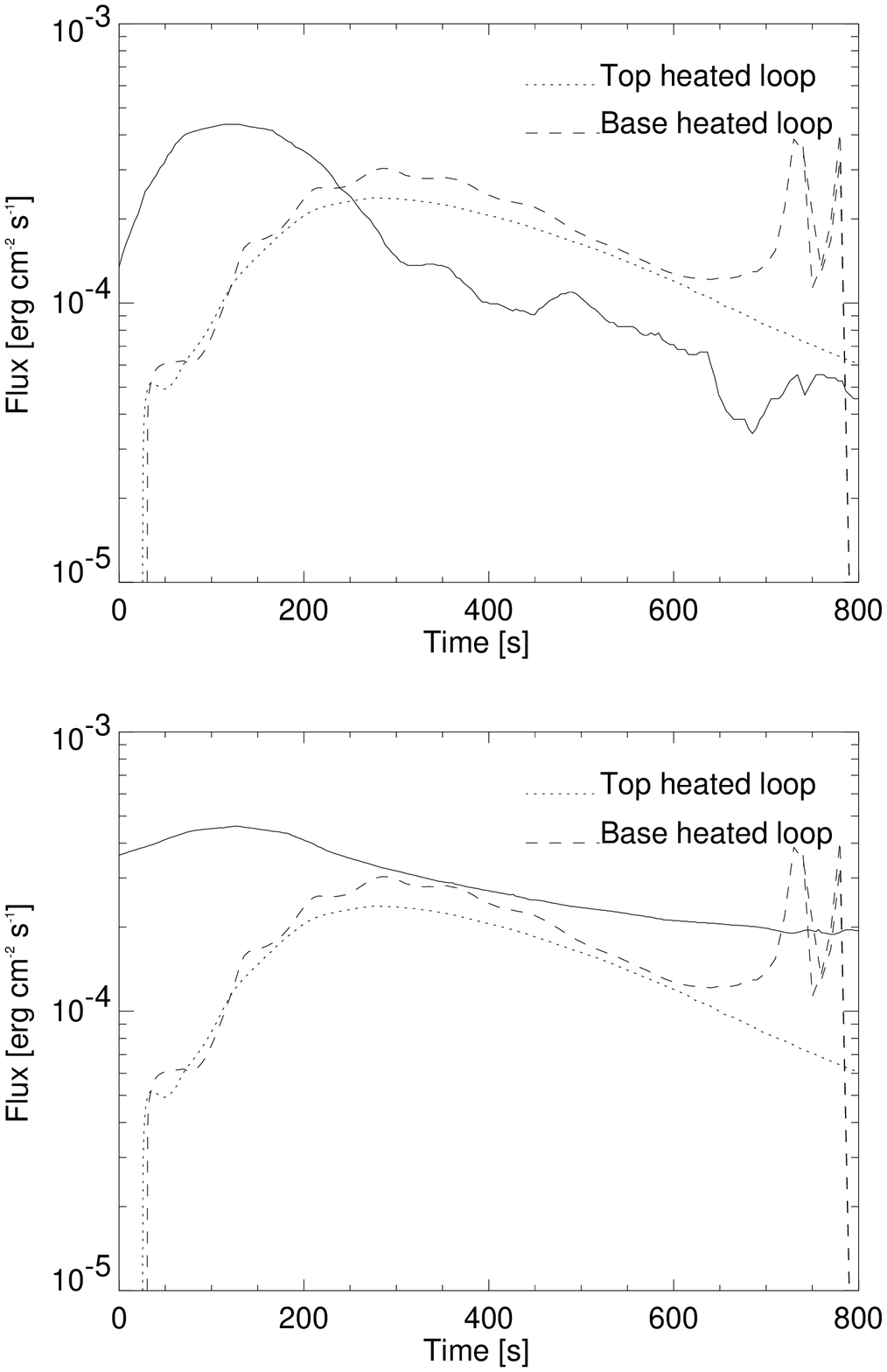,width=8.8cm}}
\caption{The emission in the \OV\ 1371.3 \AA\ line of kernel F1
(covering one foot of the loop) is compared with the results of our two
numerical calculations and  the LMF spectral synthesis model (top).
In the plot at the bottom the same results are shown together with 
the flux emitted from the whole raster (CP88).}
\label{fig:ov_LMF} 
\end{figure}

\begin{figure}[tbp]
\vspace{0.5cm}
\centerline{\psfig{figure=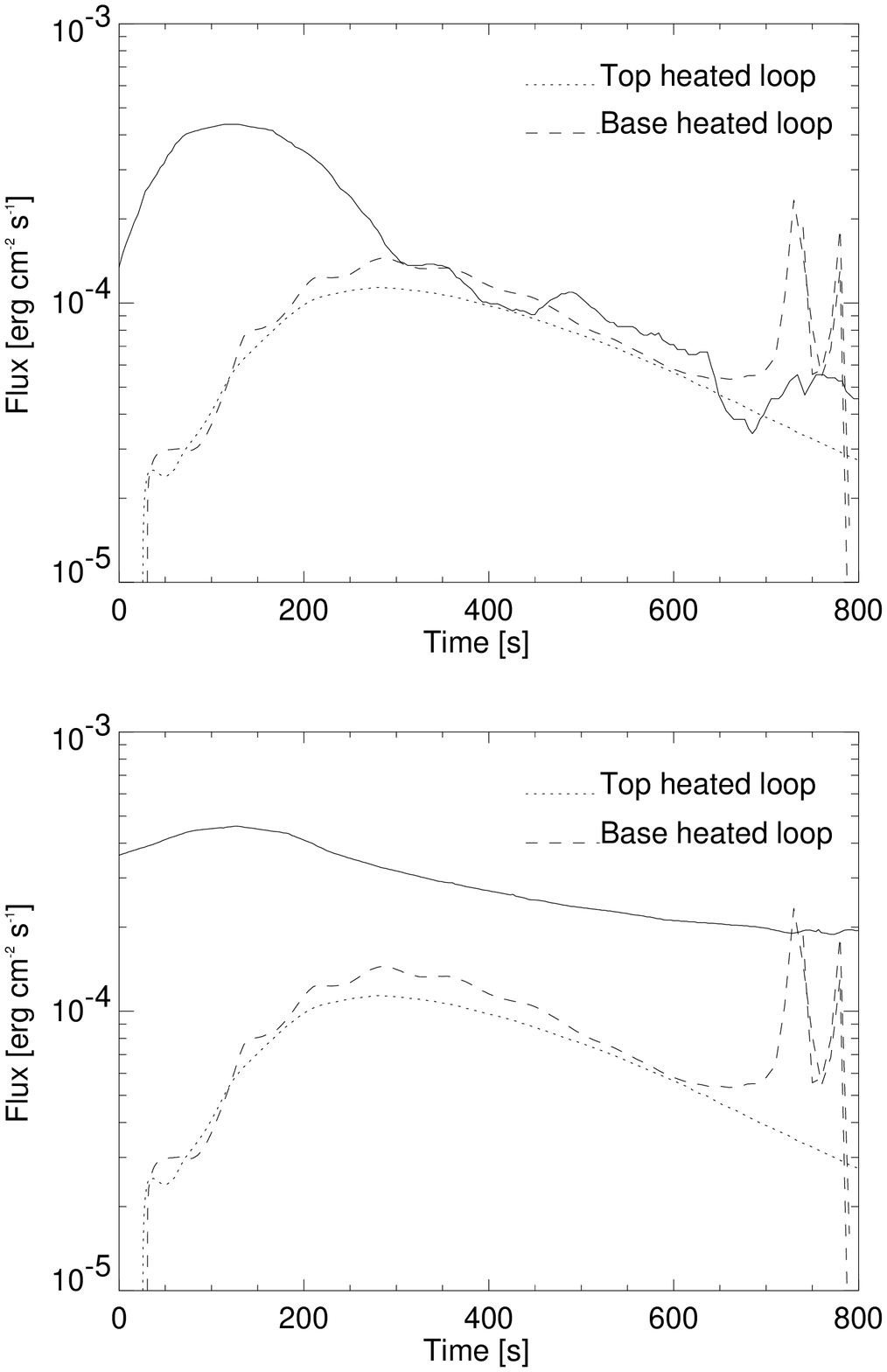,width=8.8cm}} 
\caption{The emission of kernel F1 (covering one foot of the loop) 
is compared with the results of our two numerical calculations and the
spectral synthesis model RS (top).
In the plot at the bottom the flux emitted from the whole raster (CP88)
is compared with the same modeled curves.
}
\label{fig:ov_ray} 
\end{figure}

The light curves in the EUV lines recorded during this flare have been
published in CP88.
Cheng and Pallavicini distinguished three different zones in the flaring
region, named as kernels F1, F2 and L1. The two kernels F1 and F2 covered
the two foot-points of the loop while kernel L1 corresponded to the
region between them, covering the top of the loop.
The light curves in both lines were analyzed separately for
each kernel as well as for the whole raster containing the entire flaring
region, and are shown in Figs. \ref{fig:l1fe}-\ref{fig:ov_ray}.

From the model results we calculate the flux in a particular line at a
distance $R$ from the Sun as
\begin{equation}
  F= \frac{2 A}{4\pi R^2} \int_{s_1}^{s_2} n n_e G(T)ds,  \label{eq:flux}
\end{equation} 
where $R$ is set equal to 1AU in our calculations, $A$ is the loop
cross-section (assumed constant), $G(T)$ is the emission function
tabulated by several authors for most observed lines, $n$ is the hydrogen
number density, $n_e$ the electron number density
and $s_1$ and $s_2$ are the
extreme loop points considered in the spatial integration;
we multiply the flux times 2 because the loop is supposed to be symmetric
and therefore the integral is evaluated on half a loop length.

As a first step the integral [\ref{eq:flux}] 
is calculated only in the loop volume corresponding to region L1
(two pixels).
From Fig. 10 in CP88 we can guess that the loop was along
the diagonal of kernel L1; 
assuming a semicircular geometry for the flaring coronal loop, and
considering projection effects, the flux [\ref{eq:flux}] is evaluated
from $s_1=11~ \mathrm{Mm}$ to $s_2=20~ \mathrm{Mm}$ along half a
loop, i.e. over a distance $9~ \mathrm{Mm}$ from the loop top.

Results for he \FeXXI\ 1354.1 \AA\ are shown in Figs. \ref{fig:l1fe} and
\ref{fig:rastfe} and are obtained using
the tables for the emission function $G(T)$ calculated by 
Monsignori Fossi (1994, private communication). 
The observed light curve appears similar to those obtained for the
X-ray lines with the FCS. This fact is not
surprising as the FCS lines and the \FeXXI\ 1354.1 \AA\ line originate 
in the same part of the atmosphere.

The curve computed with simulation B (base heating) is
closer to the observed one during the flare onset, and the shape
of the light curve is well simulated for the first 400s. However
the flux decreases too rapidly during the flare
decay.
On the other hand, with the heating at the top (simulation A), the
curve is too low during
the early phase, but it is high even well after the heating is
switched off. The modeled light curve peaks later than the observed one,
but follows the flare decay emission better, even though also model A
predicts a more rapid decrease of the intensity than observed,
implying that further heating might be released during the flare decay
in the EUV emitting plasma.

In Fig. \ref{fig:rastfe} the emission integrated over the whole loop
is compared with the flux in the \FeXXI\ 1354.1 \AA\ line
coming from the whole raster for both the
simulations. The modeled curves are now very different from the observed
one, as the total measured intensity is much higher than the calculated
one in both cases. 

The observed emission in the \OV\ 1371.3 \AA\ line was brighter
in both kernels F1 and F2, corresponding to the two foot-points of
the loop, than elsewhere in the raster region.
A chromospheric filament was probably responsible of the initially
increased emission in kernel F2 during the pre-flare phase (see above). The
observed flux in region F2 is higher than in F1 and the two light
curves are different (see CP88, Fig. 10), suggesting 
the presence of other emitting chromospheric structures in
region F2. For this reason we have compared our calculations
only with the data of kernel F1 and not with those of F2.
In kernel F1 there is high pre-flare emission, which could be related to
the presence of other closed structures emitting in the \OV\ 1371.3 \AA\
line or in the blended unidentified line mentioned above.

From our hydrodynamic calculations we simulate the light curve in
this line using two different formulations for the emissivity $G(T)$:
\begin{enumerate} 
\item the spectral emission function tabulated by \cite{LMF90} and
      following upgrades (thereafter LMF); 
\item the emissivity computed with Raymond's spectral code (\cite{RS77},
      and following upgrades, henceforth RS).
\end{enumerate}
The two modeled light curves have the same profile but
they differ by almost one order of magnitude.   
Discrepancies in the results obtained with different spectral models have
already pointed out by \cite{Pallavicini94} and constitute an important
difficulty in comparing real data with the hydrodynamic calculations or
any other modeling results.
Anyway, the discrepancy among spectral models does not alter significantly
the shape of the light curve, but mostly the intensity.

We have first compared the flux integrated over half a loop with the
emission from kernel F1; as a matter of fact, according to our model,
only the TR contributes to the emission in the \OV\
1371.3 \AA\ line, and therefore, only the foot-points of the loop are
bright in this wavelength.
In Fig. \ref{fig:ov_LMF} and in Fig. \ref{fig:ov_ray}
we compare our numerical results (both simulations A and B) 
with the observed flux in kernel F1 (top panel).  
We also plot the light curves predicted by our hydrodynamic calculations
and the measured flux coming from the whole raster (bottom panel).
The raster flux is larger than that predicted with
a single loop model, either if we assume the LMF spectral model or the
RS model. This fact is not surprising because many other structures
present in the raster might contribute to the emission, as for the 
\FeXXI\ 1354.1 \AA\ line.
Results obtained with the RS spectral model are in better agreement
with the observations in kernel F1.

The \OV\ 1371.3 \AA\ line peaks earlier than the \FeXXI\ 1354.1 \AA\
line in our simulations as well as in the real observation 
but the \OV\ 1371.3 \AA\ light curves of our models are qualitatively
different from the data plotted in CP88, at least for the first 5 minutes.
The model with the heating at the loop top and the RS spectral model
approximate the decay light curve quite well.
This could be said also for the bottom heated model for 
$t<700 \mathrm{s}$. Then a thermal instability occurs
(see long-dashed lines in Fig. \ref{fig:ov_LMF} and in Fig. \ref{fig:ov_ray}). 
The rapid cooling of the atmosphere, mostly due to the increase 
of radiative losses as the temperature decreases 
below $10^6 \mathrm{K}$, 
determines a significant peak of the emission which makes model B
unlikely to describe this flare.

\section{Discussion and conclusions}

The X-ray emission of this flaring loop has been accurately modeled by
\cite{P87} using the hydrodynamic results of the PH code: in that work
they obtained theoretical light curves very close to the observed
ones.  In this paper we compare the results of our numerical
calculations with both typical TR emission and coronal ones,
simultaneously.

From the analysis of all the modeled light curves in both the X-rays and EUV
bands and the data registered during the event, we note that the top
heated loop model gives results in better agreement with the coronal
emission than the bottom heated loop model, during the entire flare
evolution, in agreement with \cite{P87}. On the basis of our new
calculations we can add that this model also accounts for the observed
\FeXXI\ 1354.1 \AA\ line during a large fraction of the November 12,
1980 flare.  We note that the computed light curves in the
\FeXXI\ 1354.1 \AA\ line have a shape similar to those of the
X-ray lines formed in the same temperature range, and therefore
originating from the same part of the atmosphere.

Both our models predict \FeXXI\ line flux close to the observed one at the
time of the flare X-ray line peak, which occurred during the decay of the
flare light curve in the observed \OV\ 1371.3 \AA\ line.  Also the top-heated
model yields an already decaying OV light curve at the time of coronal
lines flare maximum.  However during the early flare the lower TR emission
cannot be fitted with the same model which consistently yields lower
emission than observed. This suggests that more plasma than modeled might
have contributed to the \OV\ 1371.3 \AA\ line emission.

In summary, the analysis of the hot coronal EUV emission (\FeXXI\ 1354.1 \AA\ line)
confirms the scenario derived with the previous study (\cite{P87})
based on X-ray lines.  The study of the \OV\ 1371.3 \AA\ line, instead,
shows a rather different evolution and the relevant light curve is
poorly fitted by the hydrodynamic loop model which instead works well
for the hot coronal emission.  We therefore confirm that the single
loop model (albeit a schematic approximation of reality) works well for
the corona of this flare, and therefore that a single loop - or a
bundle of loops evolving coherently - dominated the coronal emission,
as \cite{MacNeiceetal85} and \cite{P87} suggested.  The colder plasma
emission, instead, appears dominated by other structures, probably some
already present at an earlier phase, others evolving and reaching their
maximum emission before the \FeXXI\ 1354.1 \AA\ line peak.  Also
\cite{Feldman83}, \cite{WidingCook87}, \cite{Doschek97}, \cite{Mason98}
observed that there is a mismatch between TR and coronal flare
emission.

It appears that a single loop model is not adequate to fit the lower TR
lines simultaneously to the coronal lines (which are well fitted), at
least for this flare. To this end more complex models may be required,
such as multi-structure models.

\begin{acknowledgements}
This work has been done with partial support of the 
Italian {\em Ministero dell'Universit\`a e della ricerca Scientifica e
Tecnologica} and of the {\em Agenzia Spaziale Italiana}.
\end{acknowledgements}

\end{document}